\documentclass[%
preprint,
 aps,
]{revtex4-1}

\usepackage{wrapfig}
\usepackage{graphicx}
\graphicspath{ {./Figures/} }
\usepackage[rightcaption]{sidecap}

\usepackage{dcolumn}
\usepackage{bm}
\usepackage{hyperref}
\usepackage[mathlines]{lineno}

\newcommand{\beginsupplement}{%
        \setcounter{table}{0}
        \renewcommand{\thetable}{S\arabic{table}}%
        \setcounter{figure}{0}
        \renewcommand{\thefigure}{S\arabic{figure}}%
     }

\begin{document}


\title{Correlation-Driven Electron-Hole Asymmetry in Graphene Field Effect Devices}

\author{Nicholas Dale$^{1,2}$}
\author{Ryo Mori$^{1,3}$}
\author{M. Iqbal Bakti Utama$^{2,3,4}$}
\author{Jonathan D. Denlinger$^5$}
\author{Conrad Stansbury$^{1,2}$}
\author{Claudia G. Fatuzzo$^{1,2} $}
\altaffiliation[Current address: ]{Istituto di Scienze del Patrimonio Culturale, Consiglio Nazionale delle Ricerche (ISPC-CNR), Via Biblioteca 4, 95124 Catania, Italy}

\author{Sihan Zhao$^{1}$}
\author{Kyunghoon Lee$^{1,2}$}
\author {Takashi Taniguchi$^{6}$}
\author {Kenji Watanabe$^{7}$}
\author {Chris Jozwiak $^{5}$}
\author {Aaron Bostwick$^{5}$}
\author {Eli Rotenberg $^{5}$}
\author {Roland J. Koch$^{5}$}
\author {Feng Wang$^{1,2,8}$}
\author{Alessandra Lanzara$^{1,2,8}$}
\email{Alanzara@lbl.gov}

\affiliation{$^1$Department of Physics, University of California, Berkeley, CA, 94720, USA}
\affiliation{$^2$Materials Sciences Division, Lawrence Berkeley National Laboratory, Berkeley, CA, 94720, USA}
\affiliation{$^3$Graduate Group in Applied Science \& Technology, University of California, Berkeley, CA, 94720, USA}
\affiliation{$^4$Department of Materials Science and Engineering, University of California at Berkeley, Berkeley, CA, USA}
\affiliation{$^5$Advanced Light Source, Lawrence Berkeley National Laboratory, Berkeley, CA, 94720, USA}%
\affiliation{$^6$International Center for Materials Nanoarchitectonics, National Institute for Materials Science,  1-1 Namiki, Tsukuba 305-0044, Japan}%
\affiliation{$^7$Research Center for Functional Materials, National Institute for Materials Science, 1-1 Namiki, Tsukuba 305-0044, Japan}%
\affiliation{$^8$Kavli Energy NanoSciences Institute at University of California Berkeley and Lawrence Berkeley National Laboratory, Berkeley, CA, USA}
\date{\today}

\begin{abstract}

Electron-hole asymmetry is a fundamental property in solids that can determine the nature of quantum phase transitions and the regime of operation for devices. The observation of electron-hole asymmetry in graphene and recently in the phase diagram of bilayer graphene has spurred interest into whether it stems from disorder or from fundamental interactions such as correlations. Here, we report an effective new way to access electron-hole asymmetry in 2D materials by directly measuring the quasiparticle self-energy in graphene/Boron Nitride field effect devices. As the chemical potential moves from the hole to the electron doped side, we see an increased strength of electronic correlations manifested by an increase in the band velocity and inverse quasiparticle lifetime. 
These results suggest that electronic correlations play an intrinsic role in driving electron hole asymmetry in graphene and provide a new insight for asymmetries in more strongly correlated materials.

\end{abstract}

\pacs{Valid PACS appear here}
\maketitle
\section{\label{sec:intro}Introduction}

Electron-hole asymmetry, or the difference in a material's electronic properties upon doping with electrons versus holes, profoundly impacts the character of phase transitions\cite{Sharpe2019EmergentGrapheneb}\cite{Sarkar2020FerromagneticSuperconductor}\cite{Sajadi2018Gate-inducedInsulator}\cite{Hsu2017TopologicalDichalcogenides}, and the choice of doping for devices\cite{Arora1982ElectronTemperature}\cite{Larbalestier2001High-TcApplications}. While it typically arises from differing structures of bands containing electrons and holes\cite{Ashcroft1976SolidPhysics}\cite{Jost2017Electron-holeHgTe}, in some cases this asymmetry manifests from external sources such as impurities\cite{Yazdani1997ProbingSuperconductivity}\cite{Novikov2007NumbersGraphene}, strain\cite{Bai2015DetectingSpectroscopy}\cite{Jost2017Electron-holeHgTe}\cite{Dasilva2015TransportNitride}, or simply from intrinsic many body interactions\cite{Anderson2006TheorySuperconductors}\cite{Kretinin2013QuantumGraphene}. Graphene is an interesting case in this light, because its K point band structure is expected to be perfectly electron-hole symmetric\cite{Neto2007TheGraphene}, but the combination of its dimensionality and dispersion relation renders it highly susceptible to symmetry-breaking perturbations\cite{Kotov2012Electron-electronPerspectives}. Most experimental realizations of the monolayer\cite{Deacon2007CyclotronMonolayers}
\cite{Kretinin2013QuantumGraphene} and bilayer\cite{Zou2011EffectiveInteraction}\cite{Cao2018CorrelatedSuperlattices}\cite{Lu2019SuperconductorsGraphene} exhibit electron hole asymmetry, even after vast improvements in sample preparation, which reduce the effective strain and impurity concentration\cite{Wang2013One-DimensionalMaterial}\cite{Kretinin2013QuantumGraphene}.  Whether external sources or intrinsic interactions such as correlations\cite{Zou2011EffectiveInteraction}\cite{Anderson2006TheorySuperconductors}\cite{Cai2016VisualizingCuprates}\cite{Kretinin2013QuantumGraphene} drive asymmetry remains to be verified. This has become even more important with the recent discovery of Mott-like physics and superconductivity in bilayer graphene\cite{Cao2018CorrelatedSuperlattices}\cite{Lu2019SuperconductorsGraphene}, displaying a phase diagram which is strongly electron-hole asymmetric and reminiscent of the cuprates \cite{Gooding1994TheoryPlanes}.

The difficulty in addressing the origin of electron hole asymmetry in graphene today is the requirement of a probe that has complete access to the material self energy in both energy and momentum spanning over a large range of electron and hole dopings. Some probes, including transport \cite{Deacon2007CyclotronMonolayers} and quantum capacitance \cite{Kretinin2013QuantumGraphene}, can easily cover the broad doping range via electrostatic gating, but are only sensitive to the electronic states at the Fermi energy ($E_F$) and do not provide any momentum information. In contrast, Angle Resolved Photoemission Spectroscopy (ARPES) can provide access to the full quasiparticle spectral function $A(k,\omega)$, but so far has resorted to methods of doping that modify the fundamental properties of the system, including screening\cite{Hwang2012FermiModification}\cite{Siegel2011Many-bodyGraphene} and impurity concentration \cite{Chen2008Charged-impurityGraphene}\cite{Siegel2013Charge-carrierGraphene}. The very recent introduction of electrostatic gating into ARPES experiments \cite{Nguyen2019VisualizingHeterostructuresb}\cite{Joucken2019ASAPSpectroscopy} enables studies of the doping dependent self-energy with full energy and momentum resolution while leaving the sample in pristine condition. Here we report the first of such study, revealing significant asymmetries in the quasiparticle self energy. The doping and momentum resolution of our measurement enables us to effectively characterize the relationship between electronic correlations and these asymmetries.


 
 

\section{\label{sec:results}Results}
\begin{figure}[h]
    \centering
    \includegraphics[width=1\textwidth]{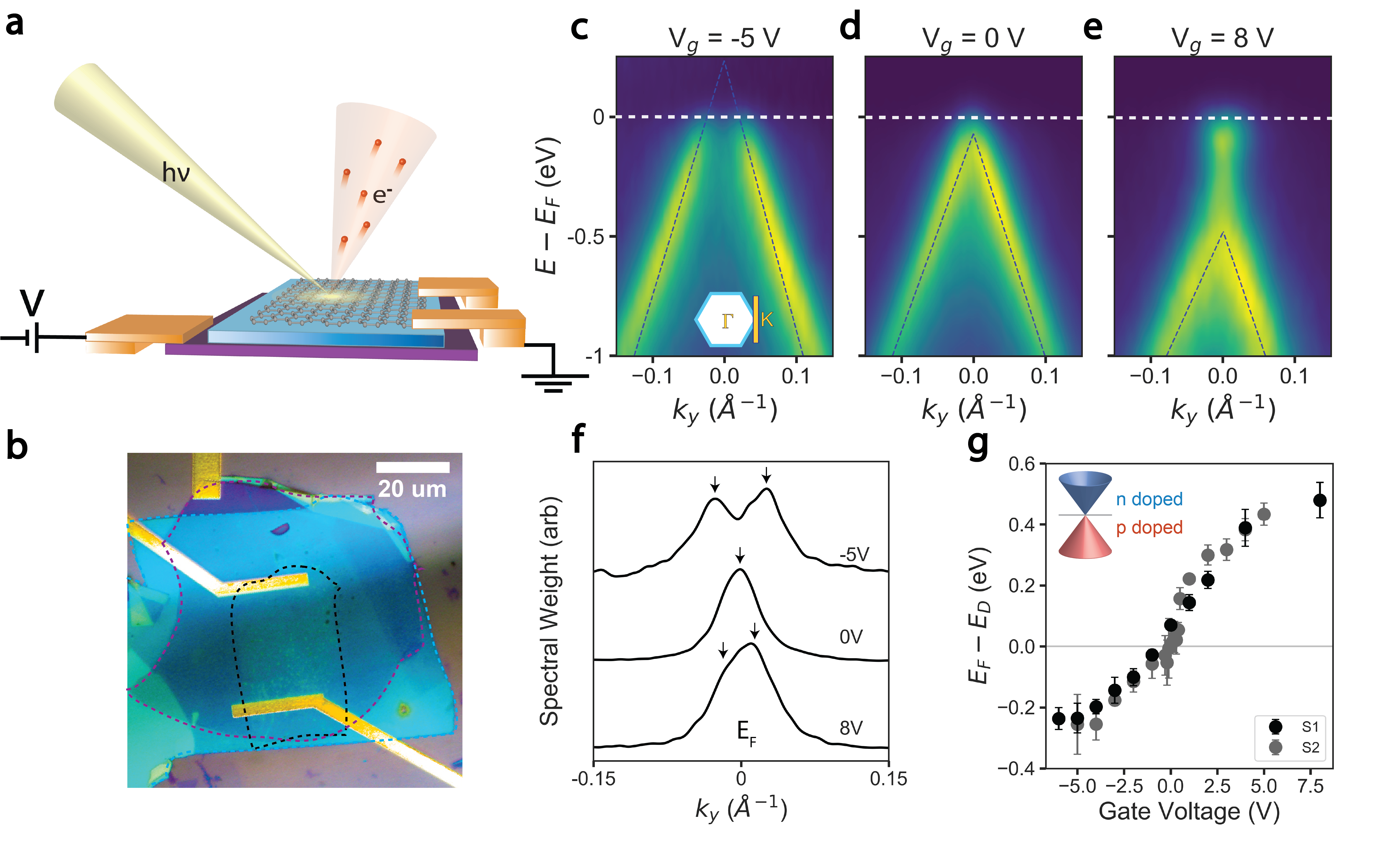}
    \caption{(\textbf{a.,b.}) schematic of experimental setup (\textbf{a-}) and optical micrograph of the graphene/hBN sample (\textbf{b-}). Dashed lines outline regions of graphene (black), hBN (blue), and graphite (purple) (\textbf{c.-e.}) ARPES spectra for S1 along the $K-K'$ direction (perpendicular  $\Gamma - K$)  at (\textbf{c-}) p doping (-5V), (\textbf{d-}) equilibrium (0V), and (\textbf{e-}) n doping (8V). Blue dashed lines indicate quasiparticle peak positions extracted from Lorentzian fits. 
    (\textbf{f.}) doping dependent MDCs spectra and quasiparticle peak positions (black arrows) at $E_F$, indicated by the white dashed line in (b-d). (\textbf{g.})  ${E_F - E_D}$ as a function of gate voltage, extracted from linear fits to the graphene spectra. Data for two different samples are shown (S1 and S2).} 
    \label{fig:1}
\end{figure}

Fig. \ref{fig:1}a presents an illustration of the sample geometry used for the ARPES experiment and gating configuration, while panel b shows the optical micrograph of the overall sample. The dashed contours identify regions of monolayer graphene (black), hBN (blue) and graphite (purple) while the yellow thick lines indicate the electrical contacts. The size of the sample is smaller than 1200 $\mu\textnormal{m}^{2}$.  The adopted beam size was 1 $\mu\textnormal{m}$ to allow 
measurements of each individual part of the sample and disentangle different contributions. The equilibrium spectra for the sample in Fig \ref{fig:1}d clearly depicts the characteristic linear bands of graphene's Dirac fermions along the $K-K'$ direction populated up to near the charge neutrality point. A positive (negative) voltage established between the graphite back gate and the graphene sample results in the addition, panel e (subtraction, panel c) of electrons to (from) the sample. Since the Fermi energy $E_F$ is held at ground, the additional negative (positive) charges shift the Dirac spectrum downward (upward). The doping change can be estimated by the peak separation at $E_F$ from momentum distribution curves (MDCs), spectra at constant energy as a function of momentum, shown in panel f for different gating values. At $V_g\,=\,0V$ the Fermi surface is a point and the momentum separation between MDC peaks is negligibly small. As electrons (holes) are added to the system, two peaks emerge and the momentum separation increases, with a maximum at $V_g\,=\,-5V (8V)$ corresponding to a p (n) doping of $2.2 \pm0.3 \cdot 10^{12} \textnormal{cm}^{-2}\, (0.6 \pm 0.3 \cdot 10^{12} \textnormal{cm}^{-2})$. The position of the Fermi energy $E_F \, - \, E_D$, displayed in Fig \ref{fig:1}g, is estimated by the intersection point of linear fits to the Dirac spectra (blue dashed lines in Figs \ref{fig:1}c-e). The scaling of $E_D$ with gate voltage away from neutrality follows $E_D \sim \sqrt{V_g}$, suggesting that the charge doping is well controlled, and a result of geometric capacitance \cite{Yu2013InteractionCapacitance}. 

\begin{figure}[!htbp]
    \centering
    \includegraphics[width=1\textwidth]{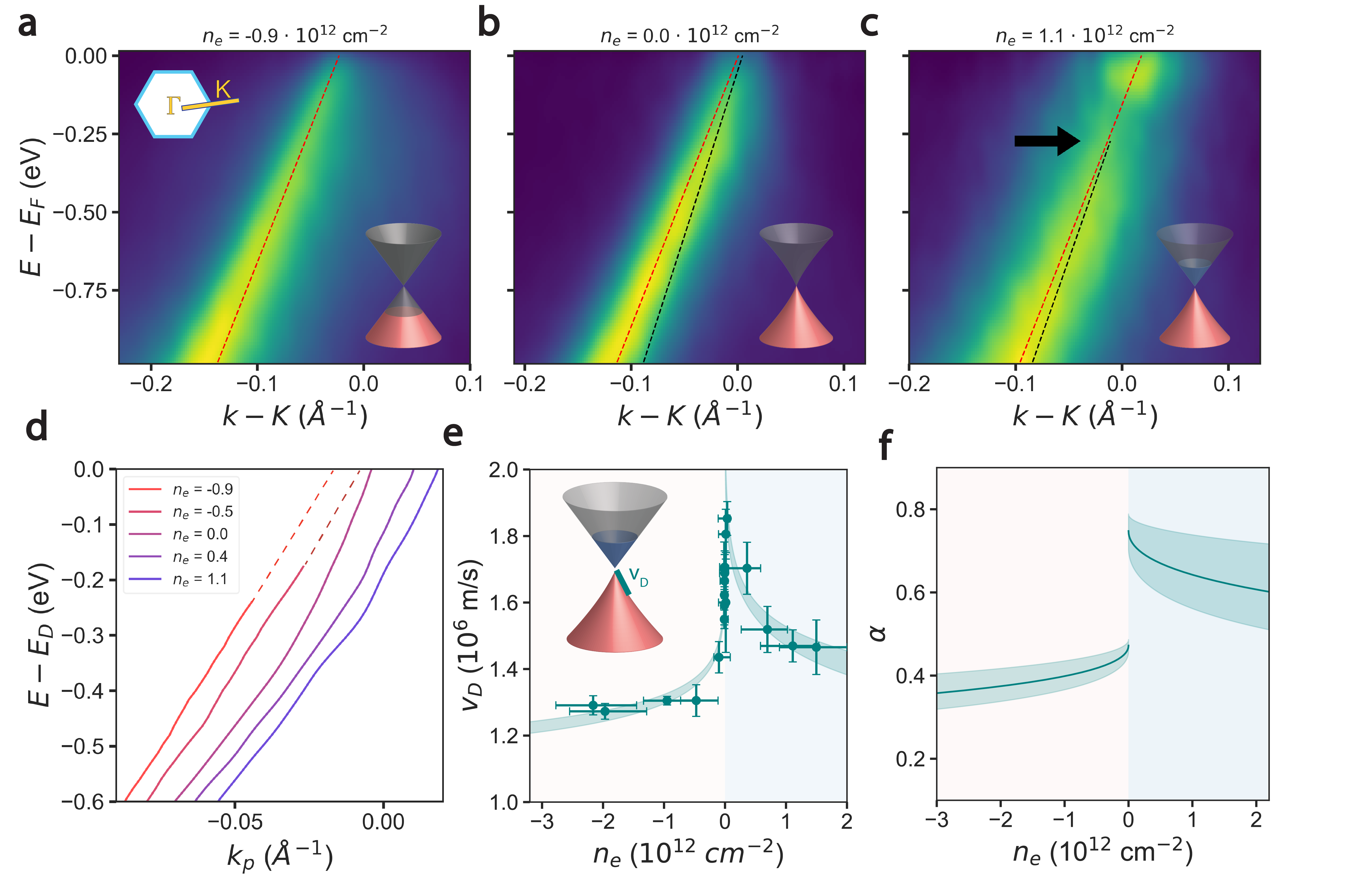}
    \caption{ (\textbf{a.-c.}) graphene spectra at three representative dopings: (\textbf{a.}) $-0.9\cdot10^{12} \textnormal{cm}^{-2}$, (\textbf{b.}) $0.0 \cdot10^{12} \textnormal{cm}^{-2}$, (\textbf{c.}) $1.1\cdot10^{12} \textnormal{cm}^{-2}$. Black (red) dashed lines indicate linear fits to dispersions near $E_D$ ($E_F$ in (\textbf{a})). Inset cartoons illustrate the deviations from Dirac cone dispersions at respective dopings. (\textbf{d}.) $E_D$ dispersions near the charge neutrality point indicate asymmetry in band velocity for electron and hole dopings. (\textbf{e}.) Extracted band velocities as a function of doping. Grey (teal) shaded regions indicate $1 \sigma$ deviation of best fit to the $v_F$ ($v_D)$ data using the logarithm-based lineshape described in the text. (\textbf{f.}) Graphene fine structure constant ($\alpha$) as a function of doping, is extracted from band velocity fits.
    }
    \label{fig:rese}
\end{figure}

Fig. 2 reports the detailed evolution of the K point electronic structure near $E_F$ for different doping (gating) values. Figs \ref{fig:rese}a-c display raw image plots near the K point for dopings of $-0.9\cdot 10^{12} \textnormal{cm}^{-2}$, $0.0\cdot 10^{12} \textnormal{cm}^{-2}$, and $1.1\cdot 10^{12} \textnormal{cm}^{-2}$ (details on calculation of the carrier density can be found in Supplementary Note 1). Already from the raw data one can see that the spectrum in Fig \ref{fig:rese}a is linear, and at the neutrality point (Fig \ref{fig:rese}b) the dispersion looks noticeably steeper near $E_F$ (= $E_D$) than at higher binding energies, in agreement with previous reports \cite{Siegel2011Many-bodyGraphene}\cite{Hwang2012FermiModification}. The electron-doped spectrum (Fig \ref{fig:rese}c) presents different structure for the valence band than does the spectrum at similar hole doping: the valence band near the Dirac point (black dashed line) is steeper than the valence band in Fig \ref{fig:rese}a (red dashed line).

These differences are better visualized by plotting the energy dispersion vs momentum (Fig \ref{fig:rese}d), extracted by fitting the momentum distribution curves with standard Lorentzian-like functions in the proximity of the Dirac point. A clear departure from linearity is observed in the data starting at the neutrality point, where the dispersion is steepest, and still observed in the electron doped side. Band velocities can be directly extracted from these data, being proportional to the slope of the ARPES dispersions. Because the dispersions for hole dopings remain linear, the band velocity at the Dirac point $v_D$ (which is above $E_F$ at these dopings) can be approximated by the Fermi velocity $v_F$. In contrast, the dispersions at neutrality (purple) and electron dopings (blue) show a large deviation from linearity, with $v_D$ nearly twice as large as velocities at $E_D - 0.5$ eV. These results clearly indicate the presence of a distinct electron-hole asymmetry in the electronic response and are summarized in panel e, where the band velocities at the Dirac point ($v_D$), extracted from the slope of ARPES dispersions, are plotted as a function of doping. Although a divergence of $v_D$ is observed in the proximity of the charge neutrality point, as previously reported \cite{Siegel2013Charge-carrierGraphene} for the electron doped side, a clear asymmetry is revealed over the entire doping range, with $v_D$ $\sim$30\% higher for electron dopings than for hole dopings. The large renormalization of the Dirac spectra was previously reported at the neutrality point and assigned to electron-electron interactions \cite{Gonzalez1994Non-FermiApproach}\cite{Siegel2013Charge-carrierGraphene} leading to a logarithmic correction of the band velocity. Using a similar model \cite{Siegel2013Charge-carrierGraphene}

\begin{equation}
\frac{v}{v_0} = 1  +\frac{\alpha}{8} \log{(\frac{n_0}{n_e})}
\end{equation}

we extract the long-range coulomb coupling $\alpha = \frac{e^2}{\epsilon \hbar v_0}$ (panel f), which is the primary contributor to the velocity enhancement. The dielectric strength $\epsilon = \epsilon_0 (1+a |n_e|^{1/2})$ is allowed to effectively increase as a function of doping \cite{Elias2011DiracGraphene}\cite{Yu2013InteractionCapacitance}, and $v_0 = 1.0 \cdot 10^6$ m/s is the local density approximation of the bare band velocity. The long range coupling strength $\alpha$ shows a strong asymmetry between the electron and hole side, which is the driver for the asymmetry in the band dispersion discussed in panel d. Though this result is in apparent contrast with some reports using $E_F$ sensitive probes \cite{Elias2011DiracGraphene}\cite{Yu2013InteractionCapacitance}, we note that the real coulomb interaction strength $\alpha$ can be isolated more reliably from energy states at the Dirac point\cite{Siegel2013Charge-carrierGraphene} rather than from states at $E_F$, which are affected by multiple different interactions \cite{Calandra2007Electron-phononSpectra}. 



\begin{figure}[!htbp]
    \centering
    \includegraphics[width=1\textwidth]{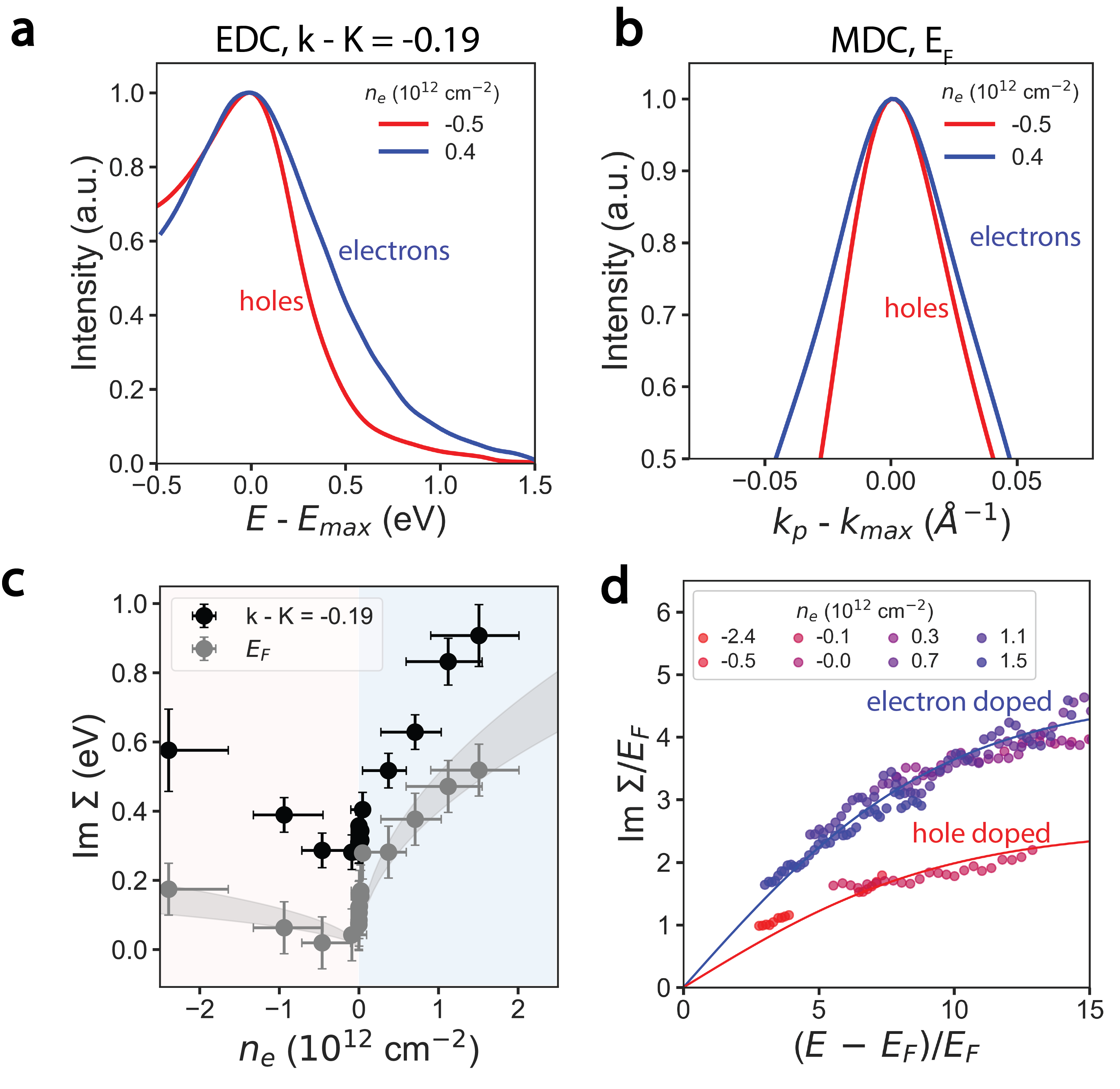}
    \caption{
   Electron-Hole Asymmetry in Self Energy. ($\textbf{a.,b.}$) Normalized EDCs at $k_p - K\, =-0.19$ ($\textbf{a-}$), and MDCs at $E_F$ ($\textbf{b-}$) for hole (red) and electron (blue) doped graphene. ($\textbf{c.}$) Imaginary part of the self energy as a function of doping for EDCs at $k_p - K\, =-0.19$ (black) and MDCs at $E_F$ (grey). Shaded grey region indicates $1 \sigma$ deviation of best fit to the square-root based function described in the text. ($\textbf{d.}$) Imaginary part of the self energy normalized by the Fermi energy. Blue (red) solid lines indicate fits to the data, as described in the text.
    }
    \label{fig:imse}
\end{figure}


Figure \ref{fig:imse} reports the imaginary part of the self energy for holes and electrons at several doping values. The MDC's FWHM $\Delta k$, the EDC's FWHM $\Delta E$ and the imaginary part of the self energy Im $\Sigma(\omega)$ are related by $2\, \textnormal{Im} \Sigma(\omega) = \hbar v_F \Delta k = \Delta E$ \cite{Valla1999Evidence}. A clear asymmetry between electrons and holes is already apparent in the raw spectra, EDC (panel a) and MDC (panel b). The full doping dependence of Im $\Sigma$ is plotted in Fig \ref{fig:imse}c for both the MDCs at $E_F$ (grey) and the EDCs at $k-K = -0.19 \AA^{-1}$ (black), each showing a strong electron hole asymmetry. Im $\Sigma$  at $E_F$ scales as $a_0 \sqrt{|n_e|}$ away from neutrality, with the amplitude $a_0 = 0.45\pm 0.05$ for electron dopings and $a_0 = 0.09\pm 0.02$ for hole dopings. This doping dependence is in contrast to alkali-doped graphene samples, which develop a $1/\sqrt{n}$ dependence from the added long range impurities\cite{Siegel2013Charge-carrierGraphene}, and the $\sqrt{n}$ scaling of the self energy at $E_F$ observed in gate-tunable graphene samples \cite{Muzzio2020Momentum-resolvedDeviceb} can be attributed to either acoustic phonons\cite{Hwang2008AcousticGraphene}\cite{Kaasbjerg2012UnravelingGraphene} or short range impurities\cite{Adam2007ATransport}. However, since both scattering sources are directly affected by the screening of the Coulomb interaction\cite{Adam2007ATransport}\cite{Hwang2008AcousticGraphene}, we can confidently conclude that the prominent electron-hole asymmetry in the self energy at $E_F$ is due to the asymmetry in the correlation strength $\alpha$ (reported in Fig \ref{fig:rese}). A similar asymmetry has been observed in a graphene sample nearly aligned with the hBN substrate \cite{Muzzio2020Momentum-resolvedDeviceb}, which may be weaker due to the competing asymmetry produced by the moir\'{e} potential\cite{Dasilva2015TransportNitride} and/or from the presence of a screening graphite back gate\cite{Stepanov2020UntyingGraphene}.

Whereas techniques that are only sensitive to the low energy physics are often marred by impurities \cite{Chen2008Charged-impurityGraphene}\cite{Deacon2007CyclotronMonolayers}, the ability of ARPES to access the entire energy range allows us to extract the intrinsic behavior of materials. Fig. \ref{fig:imse}d presents the energy dependence of the imaginary self energy scaled by the position of the Fermi energy $E_F$ for different doping values. That $\textnormal{Im} \Sigma/E_F$ collapses to two distinct curves for electron and hole dopings provides further evidence for electron hole asymmetry in the material. The reported energy dependence is qualitatively similar to numerical calculations of the inverse quasiparticle lifetime from dynamically screened electron-electron correlations\cite{Hwang2007InelasticGraphene}\cite{Tse2008BallisticGraphene}. From these calculations we can approximate the scattering rate to an empirical form:

\begin{equation}
\frac{\textnormal{Im }\Sigma_{ee}}{E_F} = c_1 \tanh{[c_2\cdot \frac{(E - E_F)}{E_F}]}
\end{equation}

where $c_1$ and $c_2$ are fit parameters. The fit shows an overall good agreement with the data and gives $c_{1h} = 2.5\pm0.2$ for hole dopings, $c_{1e} = 4.6\pm0.3$ for electron dopings, and $c_2 =0.11\pm0.03$ for both dopings. Such differences are another manifestation of the electron correlation strength, as discussed in Ref. \cite{Tse2008BallisticGraphene}.

\section{\label{sec:Discussion}Discussion}

The data reported here provide evidence of a strong electron-hole asymmetry in graphene that is driven, as we will argue below, by strong electronic correlations. We now discuss the possible sources of such asymmetry and show that it is an intrinsic property rather than driven by disorder or other extrinsic effects.

As mentioned above, there are several mechanisms that break electron-hole symmetry in graphene, and include intrinsic asymmetries in the band structure, charged impurities, and electronic correlations. The asymmetries in the band structure are induced by next nearest neighbor hopping \cite{Neto2007TheGraphene}\cite{Reich2002Tight-bindingGraphene} which can be effectively enhanced by strain\cite{Bai2015DetectingSpectroscopy}, induced for example from alignment to a substrate with a different lattice constant, and easily modeled by tight binding calculations\cite{Dasilva2015TransportNitride}. When applying the latter to our experimental data, it becomes clear that to account for the 30\% asymmetry between conduction and valence band velocity an unrealistic value of $|t'|\sim 3$ eV is needed. This is an order of magnitude larger than values reported in the literature ($t' \sim 0.3$ eV) \cite{Bostwick2007RenormalizationInteractions}\cite{Deacon2007CyclotronMonolayers}\cite{Kretinin2013QuantumGraphene}, and opposite in sign to the asymmetries produced in graphene aligned to hBN\cite{Dasilva2015TransportNitride}, and graphene strained via wrinkles\cite{Bai2015DetectingSpectroscopy}. Moreover, we note that our samples are aligned at large twist angles to the hBN substrate, where lattice reconstruction is negligibly small \cite{Jung2015OriginNitride} (see supplementary information for more details), and therefore the effect on $t'$ is negligible.

Another possible source of electron hole asymmetry is the presence of charged impurities leading, in the case of very close ($<5$ nm) proximity, to changes in the LDOS as large as 30\% \cite{Wang2013ObservingGraphene}\cite{Bai2015DetectingSpectroscopy}.
However, impurities produce an inverse quasiparticle lifetime that scales inversely with $E\, -\, E_D$ \cite{Adam2007ATransport}, in contrast to the empirical function in equation 2 used to fit our data.  Additionally, the impurity density required to produce this effect throughout a mesoscopic sample ($\sim 10^{13}\, \textnormal{cm}^{-2}$) is large enough to produce signatures in the spectral function in the form of resonance states\cite{Peres2006ElectronicCarbon}\cite{Skrypnyk2008SpectralCenters}\cite{Wang2013ObservingGraphene} or impurity bands\cite{Avsar2014Spin-orbitGraphene}\cite{Hwang2018EmergenceCerium}, which are not observed in our data. 

These observations make electronic correlations the primary driver of electron-hole asymmetry observed in our study. Indeed, this interaction can consistently explain the asymmetric logarithmic renormalization of the dispersions across charge neutrality\cite{Gonzalez1994Non-FermiApproach}\cite{DasSarma2007Many-bodyLiquid}, the nonlinear behavior of self energy at high binding energies\cite{Hwang2007InelasticGraphene}, and likely the asymmetry in the self energy at $E_F$. Finally, we note that though numerical calculations for $\Sigma_{el-el}$\cite{Tse2008BallisticGraphene} are much smaller than values found in our experiment, reaching quantitative agreement between experimental and theoretical results often requires additional scaling factors \cite{Calandra2007Electron-phononSpectra}\cite{Park2007VelocityInteraction}. 

In conclusion, we have demonstrated the power of electrostatic gated ARPES to
study the interplay of interactions and electron-hole symmetry in 2D materials. Our results point to electronic correlations as the driving force for an intrinsic electron-hole asymmetry in graphene, manifested in the dispersion and inverse quasiparticle lifetime. These findings open the intriguing possibility that electron-electron interactions might also be responsible for the asymmetries found in the phase diagrams of twisted bilayer graphene \cite{Cao2018CorrelatedSuperlattices}\cite{Lu2019SuperconductorsGraphene} and similar correlated 2D moir\'{e} systems \cite{Wang2020CorrelatedDichalcogenides}\cite{Chen2020ElectricallyGraphene}, as is found in high temperature cuprate superconductors \cite{Gooding1994TheoryPlanes}\cite{Phillips2006Mottness}\cite{Anderson2006TheorySuperconductors}\cite{Cai2016VisualizingCuprates}.
\\ \\
\textbf{Acknowledgements:}\\
We thank Salman Kahn for technical assistance in the sample fabrication setup. This work was primarily supported by the U.S. Department of Energy,
Office of Science, Office of Basic Energy Sciences, Materials Sciences and Engineering Division under Contract No. DEAC02-
05CH11231 (Ultrafast Materials Science Program KC2203). K.W. and T.T. acknowledge support from the Elemental Strategy Initiative
conducted by the MEXT, Japan ,Grant Number JPMXP0112101001,  JSPS
KAKENHI Grant Number JP20H00354 and the CREST(JPMJCR15F3), JST.
\\ \\
\textbf{Competing Interests:}\\
The authors declare that they have no competing financial interests.
\\ \\
\textbf{Correspondence:}\\
Correspondence and requests for materials should be addressed to A.L. (email: alanzara@
lbl.gov).
\\ \\
\textbf{Author Contributions:}\\
N.D. and A.L. initiated and directed the research project. T.T. and K.W. synthesized the hBN crystals. N.D., M.I.B.U., S.Z., and K.L. fabricated the graphene samples. N.D., R.M., J.D.D., C.G.F, and A.B. performed the ARPES measurements. N.D. analyzed the ARPES data using software designed by C.S., and inputs from A.L. N.D. and A.L. wrote the manuscript, with inputs from all of the authors.
\\ \\
\textbf{Data Availability:}\\The data that support the findings of this study are available from the corresponding author upon reasonable request.
\section{\label{sec:citerefs}References}
\bibliography{draft009}

\section{\label{sec:methods}Methods}
\noindent Two devices were made for this experiment -- for S1 refer to Fig \ref{fig:1} and for S2 refer to Figs \ref{fig:rese} and \ref{fig:imse} \\
\textbf{Sample Preparation:} \\
Flakes of single-layer Graphene and hexagonal Boron Nitride were exfoliated onto Silicon Wafers with 90nm-thick oxide. S1 was constructed using a method similar to that used in \cite{Utama2019VisualizationTwistb}. A stamp comprised of Polypropylene carbonate (PPC), and Polydimethylsiloxane (PDMS), and transparent tape was used to pick up Graphite, hBN, and Graphene in sequential order. The PPC stamp holding the stack was flipped onto a 90nm oxidized Si wafer with the Graphene facing up, and the polymer was subsequently removed by annealing in a vacuum furnace at 350C for 10 hours. S2 was constructed using a technique similar to that outlined in Zomer et. al \cite{Zomer2014FastNitride}. A stamp comprised of Polycarbonate (PC) and Polydimethylsiloxane (PDMS) was used to pick up the Graphene, hBN, and Graphite to form a graphene/hBN/graphite heterostructure, which was then placed onto a fresh 90nm-oxidized Si wafer. PC polymer residue was removed by placing the stack-on-chip in Chloroform for $>$60 minutes at room temperature. Contacts were patterned onto each sample surface using electron-beam lithography followed by evaporation of 5nm Cr and 50nm Au.
\\ \\  
\textbf{ARPES Measurements and Analysis:}\\
Sample 1 was measured using a Scienta R4000 Hemispherical Analyzer at the nanoARPES branch of beamline 7.0.2 (MAESTRO) at the Advanced Light Source using a photon energy of 74 eV, a temperature of 300K,and a pressure better than 1e-10 Torr. The beam was capillary refocused \cite{Koch2018NanoOptic} to a spot size of $\sim 1\mu\textnormal{m x }\,1 \mu\textnormal{m}$. The overall energy and momentum resolution was ~30meV and $0.014 \AA^{-1}$, respectively. The sample was doped electrostatically using a Keithley 2450 Source Meter.
\\
Sample 2 was measured using a Scienta R8000 Hemispherical Analyzer at Beamline 4.0.3 (MERLIN)\cite{Reininger2007MERLINALS} at the Advanced Light Source using a photon energy of 94eV, a temperature of 20K, and a pressure better than 5e-11 Torr. The beamspot was $\sim 100\mu\textnormal{m x }\,50 \mu\textnormal{m}$. The overall energy  and momentum resolution was 25meV and $0.017 \AA^{-1}$, respectively. The sample was doped electrostatically using a Keithley 2200 programmable power supply electrically connected to the cryostat.
\\
All ARPES data in this paper were analyzed using pyARPES, an open-source python-based analysis framework\cite{Stansbury2020PyARPES:Spectroscopies}. Spectra presented in the figures have had a background (estimated by mean value of detector counts $\simeq0.5\AA^{-1}$ away from the K point) removed, and are smoothed by a gaussian filter with windows in momentum and energy smaller than the experimental resolution. 

\section{\label{sec:supp}Supplementary Material}
\beginsupplement

\noindent\textbf{Supplementary Note 1: Carrier Density Measurements}
\begin{figure}[!htbp]
    \centering
    \includegraphics[width=1\textwidth]{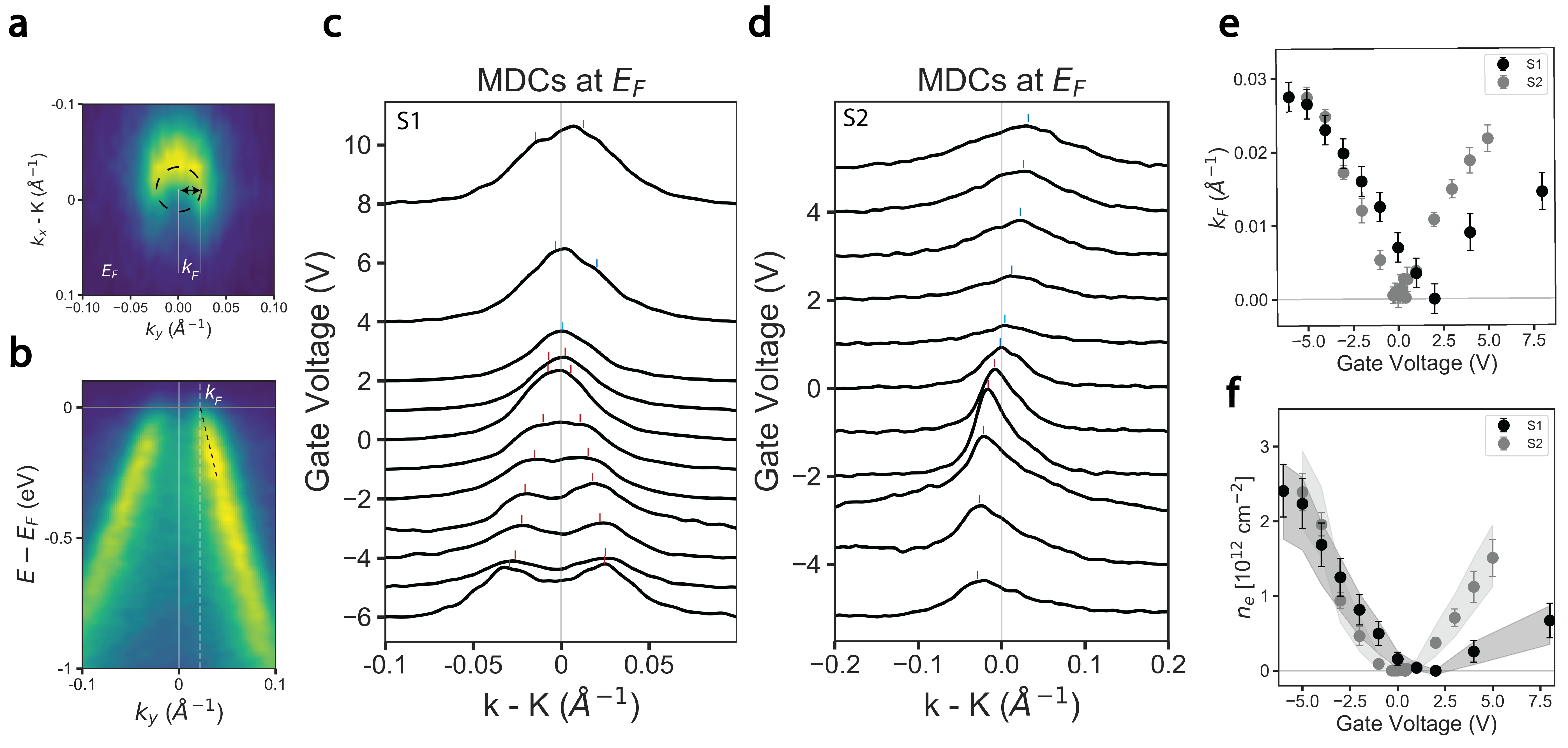}
    \caption{(\textbf{a.}) Fermi surface for S1 at $V_g = -6V$. White lines indicate Fermi wavevector $k_F$ measured as the radius of the Fermi surface. (\textbf{b.}) Graphene spectra cut along $k_x = K$, with illustration of more accurate method to extract $k_F$. The position of $k_F$ (white vertical line) is determined from intersection of band dispersion (dashed black line) with the Fermi level (horizontal grey line).(\textbf{c.,d.}) MDCs at $E_F$ as a function of applied gate voltage for S1 (\textbf{c-}) and S2 (\textbf{d-}). Valence (conduction) band positions labelled with red (blue) markers. (\textbf{e.}) Fermi wavevector $k_F$ as a function of gate voltage for both samples. Error bars are estimated from the experimental momentum resolution and the broadness of the MDC. (\textbf{f.}) Carrier density $n_e$ as a function of gate voltage for both samples, calculated using Luttinger's theorem: $n_e = k_F^2/\pi$. Error bars represent errors propagated from estimates of $k_F$, while shaded regions represent errors propagated from uncertainty in the charge neutrality point.
    } 
    \label{fig:supp-carrier}
\end{figure}

The charge-carrier density can be obtained from the size of the Fermi surface using Luttinger's theorem. Assuming a conical dispersion for graphene, the Fermi surface is a circle with a radius of the Fermi wavevector $k_F$, i.e. $n_e = k_F^2/\pi$ (see Fig \ref{fig:supp-carrier}a, and \ref{fig:supp-carrier}b). We extract the Fermi wavevector from the distance between peak positions at the Fermi level, which are displayed in Fig \ref{fig:supp-carrier}c for S1 and Fig \ref{fig:supp-carrier}d for S2. Whereas data from S1 includes contributions from both sides of the Dirac cone, data from S2 primarily contains one band, and $k_F$ can be extracted more accurately as (see Fig \ref{fig:supp-carrier}b) the intersection of the band dispersion (black dashed line) with the Fermi level (grey line). The K point position for S2 is then extracted as value of $k_F$ at the doping with the steepest $v_D$, which in our case occurs at a gate voltage of 0.5V (see Figure \ref{fig:rese}). The summary of extracted $k_F$ values, and calculated $n_e$ values are presented in Fig \ref{fig:supp-carrier}e and Fig \ref{fig:supp-carrier}f, respectively. For S1, error bars for $k_F$ are estimated based on the broadness of the bands and the momentum resolution, which was $\sim 0.014\, \AA^{-1}$, whereas for S2, error bars are extracted from the statistical error in the linear fit to the dispersion. Error bars for $n_e$ are obtained by propagating errors in $k_F$. Away from charge neutrality, both devices appear to have a roughly linear dependence between doping and gate voltage i.e. $n_e \simeq C V_g$, which is expected when treating the system as a parallel-plate capacitor with geometric capacitance $C$. The dip in geometric capacitance near the neutrality point indicates contributions of the quantum capacitance, which scale with the strength of electron-electron interactions\cite{Yu2013InteractionCapacitance}. 
\\
\\
\textbf{Supplementary Note 2: MDC and EDC Analysis}\\

\begin{figure}[!htbp]
    \centering
    \includegraphics[width=1\textwidth]{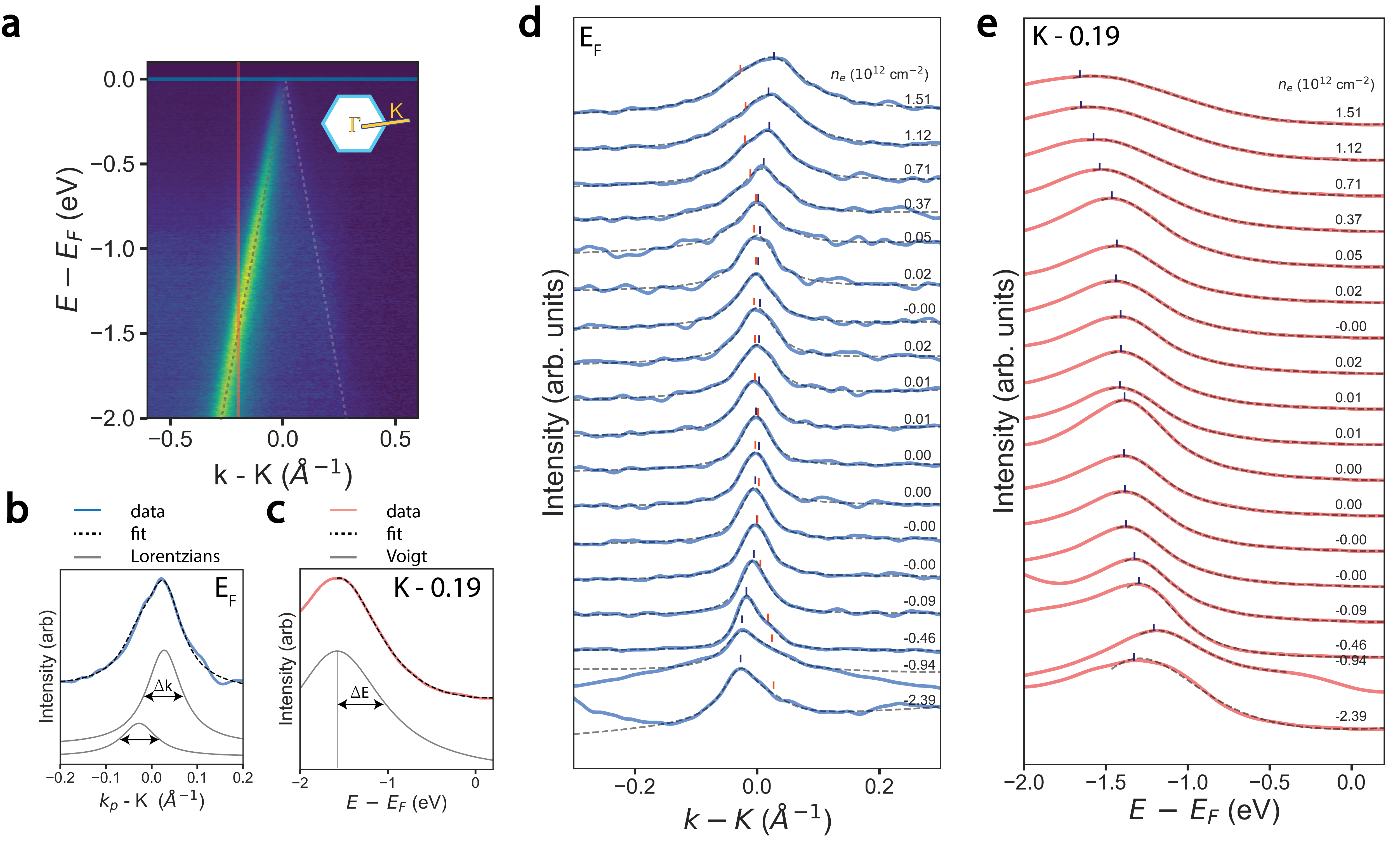}
    \caption{(\textbf{a.}) Graphene spectra for sample S2 at $V_g = 0$. Dashed lines are guides to eye for  valence bands. Light blue horizontal lines indicate binding energies at which band width $\Delta k $ is measured from MDC fits, red vertical line indicates momentum at which band width ($\Delta E$) is measured from EDC fits. (\textbf{b.}) Example of a Lorentzian lineshape fit to single MDC at $E_F$  (\textbf{c.}) Example Voigt fit to EDC at $K - 0.19 \AA^{-1}$. (\textbf{d.,e.}) MDCs at $E_F$ (\textbf{d-}) and EDCs at $K - 0.19 \AA^{-1}$ (\textbf{e-}) and peak fits (black), which contribute to figure \ref{fig:imse}c in the primary manuscript. Peak positions are indicated by navy and orange markers. Data from progressively higher dopings are offset for clarity.
    } 
    \label{fig:supp-mdc}
\end{figure}
Graphene scattering rates described in the main text are extracted using the following procedure.  The quasiparticle scattering rate $\Gamma$ is proportional to the width of MDCs $\Delta k$, which usually  have a Lorentzian lineshape \cite{Damascelli2003Angle-resolvedSuperconductors}. Fig \ref{fig:supp-mdc}a presents a typical graphene spectra for S2, which has matrix elements strongly favoring the $k - K < 0$ side of the Dirac cone. The faint presence of the other side of the Dirac cone for $k - K > 0$  requires MDCs at $E_F$ to be fit to two Lorentzians with an affine background. We restrict the two peaks to have the same width and be placed symmetrically about the Dirac point,  (see Fig \ref{fig:supp-mdc}b). EDCs at higher binding energies, for example at $K - 0.19 \AA^{-1}$ can simply be fit to a single Voigt lineshape with an affine background (see Fig \ref{fig:supp-mdc}c). Figures \ref{fig:supp-mdc}d and \ref{fig:supp-mdc}e show the MDCs (blue) and EDCs (red) as well as the fits (black) used to extract the quasiparticle scattering rate in the main text. MDCs have been smoothed using a gaussian filter with a momentum window sized smaller than the expermental resolution, and EDCs have been smoothed using a window (60 meV) that is smaller than the smallest recorded value of Im $\Sigma$ for sample S2. The statistical error of the fit is used to make the error bars in Fig \ref{fig:imse}c. The presence of hBN spectra can be seen in the MDCs at hole dopings $n_e < -0.5\cdot 10^{12} \textnormal{cm}^{-2}$, and in the EDCs at hole dopings $n_e < 0.0\cdot 10^{12} \textnormal{cm}^{-2}$, though they do not strongly affect the extracted MDC or EDC widths due to the large discrepancy in broadness of hBN and graphene bands.
\\
\\
\textbf{Supplementary Note 3: Alignment to the hBN Substrate}\\

\begin{figure}[!htbp]
    \centering
    \includegraphics[width=1\textwidth]{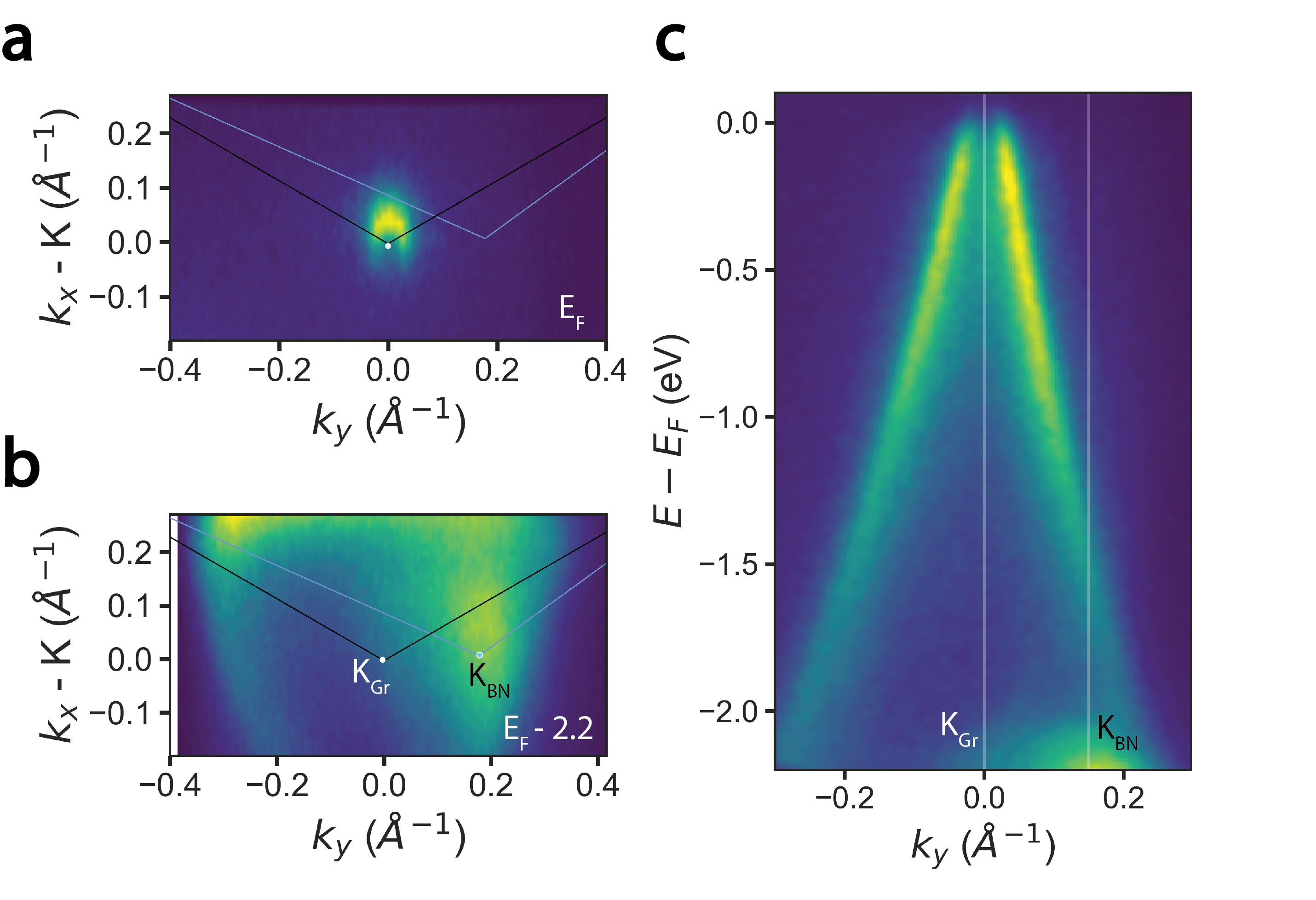}
    \caption{Constant energy maps for S1 for a gate voltage of $V_g = -6V$ at $E_F$ (\textbf{a.}) and $E_F -2.2 eV$ (\textbf{b.}). Black (blue) lines outline the graphene (hBN) Brillouin Zone, with the K point labelled by the white (blue) dot. hBN axes are rotated 6 degrees w.r.t. the Graphene ones. \textbf{c.} Photoemission spectra along $k_x = K$ indicate conical graphene bands near the Fermi level and the apex of parabolic hBN bands at $\sim$2.2 eV below $E_F$.
    } 
    \label{fig:supp-hbnalign}
\end{figure}

Alignment angle between the graphene sample and hBN substrate is extracted by overlaying their respective Brillouin Zones atop of photoemission data in Figs \ref{fig:supp-hbnalign}a, and \ref{fig:supp-hbnalign}b. The graphene K point is characterized by the radial center of the Dirac cone dispersion near $E_F$ (Fig \ref{fig:supp-hbnalign}a) and the hBN K point is characterized as the peak of a parabolloid band at a binding energy of $\sim$ 2.2 eV as seen from Fig \ref{fig:supp-hbnalign}c. The angular orientation of the hBN can also be determined using the equation $\theta \simeq 2 \arcsin(\frac{\Delta k}{2 K})$ \cite{Cao2018UnconventionalSuperlattices}. Using the momentum separation $\Delta k \simeq 0.18 \AA^{-1}$ extracted from Fig  \ref{fig:supp-hbnalign}c, we obtain an relative alignment of $\sim 6$ degrees, which agrees with the alignment of the overlaid Brillouin Zones in Figs \ref{fig:supp-hbnalign}a and \ref{fig:supp-hbnalign}b. Using the same technique, we obtain a relative twist of $\sim 13$ degrees for sample S2.
\\
\\
\textbf{Supplementary Note 4: Shift of K point upon Application of Gate Voltage}\\

\begin{figure}[!htbp]
    \centering
    \includegraphics[width=1\textwidth]{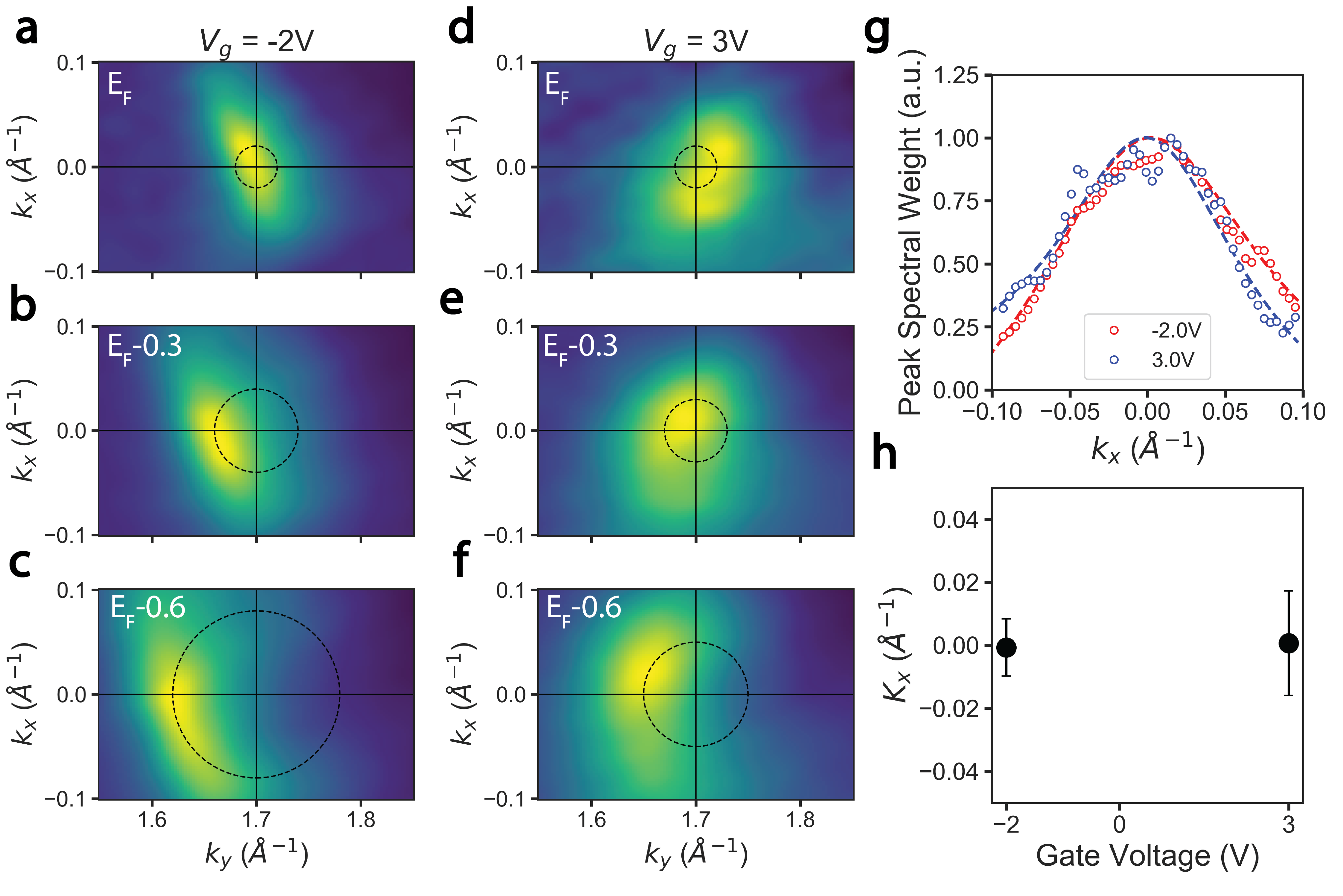}
    \caption{(\textbf{a.-c.}) Constant energy spectra for a gate voltage of $V_g\,=\,-2\textnormal{V}$ at $E_F$ (\textbf{a-}), $E_F-0.3$ eV (\textbf{b-}), and $E_F-0.6$ eV (\textbf{c-}). (\textbf{d.-f.}) Same as (\textbf{a-c}) but for $V_g\,=\,3\textnormal{V}$. Black crosshairs indicate the location of the K point, and dashed circles are guide to the eye. (\textbf{g.}) Spectral weight obtained from the peak height of MDCs at $E_F$ as a function of momentum for $V_g\,=\,-2\textnormal{V}$ (red), and $V_g\,=\,3\textnormal{V}$ (blue). (\textbf{d.}) $k_x$ values for the center of Lorentzian peaks (dashed curves) fitted to the data in the previous panel, with errors bars representative of the statistical error of the peak fit. 
    } 
    \label{fig:supp-kalign}
\end{figure}

The analysis presented here is very sensitive to the exact identification of the K point. Indeed, given the linear dispersion of graphene, a small misalignment within the sample can cause a large error in the doping value. Previous gated ARPES experiments have suffered from stray fields that drive a shift of the measured spectra with gating \cite{Muzzio2020Momentum-resolvedDeviceb}. To minimize the effects of experimental stray fields, in our experiment we have adopted a sample geometry with a sizeable sample grounding plane \cite{Nguyen2019VisualizingHeterostructuresb}. The resulting constant energy maps for our sample S2 (see panels a-f) show that the K point does not shift significantly for negative (-2V) and positive (+3V) gate voltages, allowing a precise identification of the K point from the peak in the spectral weight at $E_F$ (see panel g). With this geometry we find that the K point shift is negligible and much less than our momentum resolution ($\sim 0.017 \AA^{-1}$).

\end{document}